\documentclass[
    ,final            
    ,sort                 
  ]
  {aipproc}

\providecommand{\tabularnewline}{\\}

\layoutstyle{8x11single}

\usepackage{amsmath}
\usepackage{amstext}
\usepackage{psfrag}

\begin{document}


\title{Correlated Gaussian method for dilute bosonic systems}

\author{H.~H. Sørensen}{
  address={Department of Physics and Astronomy, University of Aarhus,
Denmark}
}

\author{D.~V. Fedorov}{
  address={Department of Physics and Astronomy, University of Aarhus,
Denmark}
}

\author{A.~S. Jensen}{
  address={Department of Physics and Astronomy, University of Aarhus,
Denmark}
}

\begin{abstract}
The weakly interacting trapped Bose gases have been customarily described
using the mean-field approximation in the form of the Gross-Pitaevskii
equation. The mean-field approximation, however, has certain limitations,
in particular it can not describe correlations between particles.
We introduce here an alternative variational approach, based on the
correlated Gaussian method, which in its simplest form is as fast
and simple as the mean-field approximation, but which allows successive
improvements of the trial wave-function by including correlations
between particles. 
\end{abstract}

\maketitle


\section{Introduction}

Dilute Bose systems trapped in external fields have been a rapidly
developing field since the Bose-Einstein condensation was observed
almost a decade ago. Theoretically the mean-field approach in the
form of the Gross-Pitaevskii equation \cite{Pitaevskii} has been widely
and successfully applied to these systems. The computational complexity of
the method, and thus the computational time, is independent of the number
of particles $N$, in other words it is of the order of $O(1)$.Therefore
the method can be applied for large (mesoscopic) bosonic systems,
especially when combined with a pseudo-potential (in the form of
the $\delta$-function potential) approximation for the
interaction potential between particles.

However, the mean-field method has certain limitations, in particular
it cannot be easily extended to include correlations
between particles. Correlations become important for systems with
higher densities and/or stronger effective interactions. Such strong
interacting regimes, where the mean-field theory breaks down \cite{Blume},
are now routinely achieved experimentally by use of Feshbach resonances.

Rigorous many-body methods, like the diffusion Monte-Carlo method
\cite{Blume,DuBois}, which include all correlations,
has computational complexity of the order $O(N^{2})$
and therefore can only be applied for smaller systems. Again, for
relatively dilute gases only few simple types of correlation
are expected to be important, and including the full machinery of rigorous
few-body methods is perhaps by far an overkill for these systems.

Recently, several methods with computational complexity $O(1)$ have
been proposed for finite-range \cite{Ole,Das}, and zero-range \cite{Sogo}
interactions, where the trial wave-function can incorporate two-body
correlations. However, these methods can not be easily extended to
include higher order correlations.

We introduce here yet another approach which has an important advantage
over the existing methods. Namely the approach can incorporate any desired
number and type of correlations -- from an uncorrelated wave-function
with computational complexity of $O(1)$ at one end, to fully correlated
wave-function with computational complexity of $O(N^2)$ and higher at the
other end. Thus, depending upon the problem at hand one has a possibility
to negotiate a reasonable trade off between the sophistication of the
trial wave-function and the computational time.

For dilute gases only few types of lowest order correlations should
be of importance, and it turns out that for these types of correlations
the method is yet of $O(1)$ order of complexity.

The method is based on correlated Gaussians and amount to a judicious
choice of the nonlinear parameters of the basis Gaussians.

\section{Method}

\subsection{Jacobi coordinates}

Consider a system of $N$ particles with masses $m_{i}$, coordinates
$\mathbf{r}_{i}$, $i=1..N$, and the Hamiltonian
\begin{equation}\label{eq:H}
H=-\sum_{i=1}^{N}\frac{\hbar^{2}}{2m_{i}}\frac{\partial^{2}}{\partial\mathbf{r}_{i}^{2}}+\sum_{i<j}V_{ij}(\mathbf{r}_{i}-\mathbf{r}_{j})+V_{ext}\;,
\end{equation}
where $V_{ij}$ is the potential between particles $i$ and $j$
and $V_{ext}$ is the external confining potential (a trap). Usually
the trap is assumed to be harmonic,
\begin{equation}
V_{ext}=\sum_{i=1}^{N}\frac{1}{2}m_{i}\omega^{2}r_{i}^{2}\;.
\end{equation}
It is of advantage to introduce the scaled coordinates, $\mathbf{q}_{i}=\sqrt{\frac{m_{i}}{m}}\mathbf{r}_{i}$,
where $m$ is an arbitrary mass scale. Indeed the kinetic energy
operator $T$ and also the harmonic trap potential $V_{ext}$ have
then a more symmetric form,
\begin{equation}\label{eq:Kq}
T=-\frac{\hbar^{2}}{2m}\sum_{i}\frac{\partial^{2}}{\partial\mathbf{q}_{i}^{2}}\;,\;
V_{ext}=\frac{1}{2}m\omega^{2}\sum_{i}\mathbf{q}_{i}^{2}.
\end{equation}
The Jacobian of the transformation from $\mathbf{r}$ to the scaled
coordinates $\mathbf{q}$ is equal to\begin{equation}
\frac{\partial(\mathbf{q}_{1}..\mathbf{q}_{N})}{\partial(\mathbf{r}_{1}..\mathbf{r}_{N})}=\prod_{i}\left(\frac{m_{i}}{m}\right)^{3/2}\;.\end{equation}
 If all particles have the same mass $m$, there is no difference
between coordinates $\mathbf{r}$ and $\mathbf{q}$.

One can make a further suitable linear transformation to a new set
of coordinates,
\begin{equation}\label{eq:tr}
\mathbf{x}_{i}=\sum_{j}U_{ij}\mathbf{q}_{j},
\end{equation}
or, in matrix notation $\mathbf{x}=U\mathbf{q}$, where the matrix $U$ is
independent of $\mathbf{q}$. The new system of coordinates is called
Jacobi if i) one of the coordinates, say the $N$th, is
proportional to the center of mass coordinate $\mathbf{R}$ of the
system, $\mathbf{x}_{N}=\sqrt{\frac{\sum_{i}m_{i}}{m}}\mathbf{R}$;
ii) the other $N-1$ coordinates are translation invariant; and iii)
the transformation preserves the ``diagonal'' form (\ref{eq:Kq})
of the kinetic energy operator.

The last property implies that the transformation (\ref{eq:tr}) and
also any transformation between different Jacobi coordinates is unitary,
$UU^{T}=1$ (where $^{T}$ denotes transposition), with the corresponding
Jacobian being equal to unity.  The unitarity means that the so-called
hyper-radius $\rho$, defined as $\rho^{2}\equiv\sum\mathbf{q}_{i}^{2}$,
is invariant under these transformations,
\begin{equation}
\rho^{2}\equiv\sum_{i}\mathbf{q}_{i}^{2}
=\sum_{i}\mathbf{x}_{i}^{2}=\frac{1}{m}\sum_{i}m_{i}\mathbf{r}_{i}^{2}\;.
\end{equation}

With Jacobi coordinates the center of mass coordinate decouples and the
hyper-radius it therefore often defined without the contribution
from the center of mass coordinate $\mathbf{x}_N$,
\begin{equation}
\rho^{2}=\sum_{i<N}\mathbf{x}_{i}^{2}
=\sum_{i}\mathbf{q}_{i}^{2}-N\mathbf{R}^{2}\;.
\end{equation}

One of the possible choices of the Jacobi coordinates is
\begin{equation}
\mathbf{x}_{i=1..N}=\sqrt{\frac{\mu_{i}}{m}}\left(\mathbf{R}_{i}-\mathbf{r}_{i+1}\right)\;,\label{eq:x}
\end{equation}
where $\mathbf{R}_{i}$ is the coordinate of the center of mass of
the first $i$ particles, $\mathbf{r}_{N+1}\equiv0$, and $\mu_{i}$
is the reduced mass
\begin{equation}
\mu_{i}=\frac{M_{i}m_{i+1}}{M_{i}+m_{i+1}}\;,
\end{equation}
where $M_{i}=\sum_{k=1}^{i}m_{k}$.

In the following we shall only consider identical particles with
$m_{i}\equiv m$.

\subsection{Hyper-radial approximation}

\subsubsection{Non-interacting bosons in a harmonic trap}

Let us consider a system of non-interacting bosons in a harmonic trap.
This should be a good first approximation to a system of weakly interacting
bosons in a trap which is smooth at the bottom and spherically symmetric.

The ground-state wave-function $\Psi$ of a system of non-interacting
bosons is a product
\begin{equation}
\Psi=\prod_{i}\psi_{0}(q_{i}),
\end{equation}
where $\psi_{0}(q)$ is the lowest ($s$-wave) single-particle state of the trap.
If the trap is harmonic, $\psi_{0}(q)$ is a Gaussian,
$\psi_{0}(q)\propto e^{-\frac{1}{2}\alpha_{0}q^{2}}$,
where $\alpha_{0}^{-1/2}$ is the (scaled) oscillator length, and
the ground-state wave-functions simplifies to a single Gaussian
depending only on the hyper-radius $\rho$,
\begin{equation}
\Psi=\prod_{i}\psi_{0}(q_{i})
\propto\prod_{i}e^{-\frac{1}{2}\alpha_{0}q_{i}^{2}}
=e^{-\frac{1}{2}\alpha_{0}\sum_{i}q_{i}^{2}}
=e^{-\frac{1}{2}\alpha_{0}\rho^{2}}.
\end{equation}

A single Gaussian $e^{-\frac{1}{2}\alpha_{0}\rho^{2}}$ is thus an exact
solution for a system of non-interacting bosons in a harmonic trap.
Generally speaking a function of hyper-radius will provide an exact
solution to the many-body system in cases where the potential energy
of the system depends only on the hyper-radius. The harmonic trap is
precisely this type of potential.

\subsubsection{Weakly interacting bosons}

If the particles in the trap interact only weakly one can assume,
following the ideas from the mean-field theory, that the inter-particle
interactions will effectively lead to a certain modification of the
field. The solution will then be some square-integrable function
of hyper-radius, $\Phi_{HR}(\rho)$, which can be represented as a
linear combination of, say, $n$ Gaussians,
\begin{equation}
\Phi_{HR}(\rho)=\sum_{s=1}^{n}C_{s}e^{-\frac{1}{2}\alpha_{s}\rho^{2}}
=\sum_{s}\prod_{i}C_{s}e^{-\frac{1}{2}\alpha_{s}q_{i}^{2}},
\end{equation}
where $C_{s}$ are variational parameters, and the range parameters
$\alpha_{s}$ ($s=1..n$) are assumed to be fixed and chosen to span
the necessary functional space.  This trial wave-function is called a
\emph{hyper-radial approximation}. In practice the parameters $\alpha_{s}$ are
chosen and then optimized in a stochastic procedure using the ideas from
the stochastic variational method \cite{SVM}.

\subsubsection{Hyper-radial vs. mean-field}

The variational mean-field approach is based on an assumption that
a product wave-function can provide a good description of an
interacting system. The trial wave-function $\Psi_{MF}$ is taken as a product
of single-particle functions $\psi$,\begin{equation}
\Psi_{MF}=\prod_{i}\psi(q_{i}),\end{equation}
 where the functional form of $\psi(q)$ is varied to reach the minimum
of the expectation value of the Hamiltonian. Assuming that $\psi$
is a square integrable function, one can represent it as a linear
combination of Gaussians,
\begin{equation}
\psi(q)=\sum_{s}c_{s}e^{-\frac{1}{2}\alpha_{s}q^{2}},
\end{equation}
where the coefficients $c_{s}$ are the variational parameters. The
trial mean-field wave-function then becomes
\begin{equation}\label{eq:MF}
\Psi_{MF}=\prod_{i}\sum_{s}c_{s}e^{-\frac{1}{2}\alpha_{s}q_{i}^{2}},
\end{equation}
which should be compared with the hyper-radial trial wave-function
\begin{equation}\label{eq:HR}
\Phi_{HR}(\rho)=\sum_{s}\prod_{i}C_{s}e^{-\frac{1}{2}\alpha_{s}q_{i}^{2}}.
\end{equation}

The two trial functions (\ref{eq:MF}) and (\ref{eq:HR}) are similar
but not equivalent since the sum and the product operators generally do
not commute. Note that the hyper-radial variational parameters
$C_{s}$ are linear, while the mean-field parameters $c_{s}$ are
non-linear\footnote{indeed the Gross-Pitaevskii mean-field equation
is non-linear.}. In practice, however, as we shall show by numerical
calculations, both trial functions give rather similar results.

Both functions are totally symmetric and thus do not require an explicit
symmetrization. The computational time for the variational minimization
of the Hamiltonian with both functions is independent of
the number of particles.

The hyper-radial function has an advantage that the center of mass
motion can be easily decoupled by a (unitary) transformation to relative
Jacobi coordinates.  Again, the mean-field function cannot be easily
improved, while the hyper-radial function is only the basis for further
improvements.

\subsection{Correlations}

\subsubsection{Two-body correlations}

The correlation between a pair
of particles can be described by a basis function in the form
\begin{equation}
\Phi_{12} =
e^{-\frac{1}{2}\alpha\rho^{2}
-\frac{1}{2}\beta(\mathbf{q}_{1}-\mathbf{q}_{2})^{2}},
\end{equation}
where there are now two independent parameters, $\alpha$ and $\beta$.
The trial wave-function is then a linear combination of $\Phi_{12}$'s
with different parameters $\alpha$ and $\beta$,
\begin{equation}
\Psi=\sum_{s,u}C_{su}e^{-\frac{1}{2}\alpha_{s}\rho^{2}
-\frac{1}{2}\beta_{u}(\mathbf{q}_{1}-\mathbf{q}_{2})^{2}},
\end{equation}
where $C_{su}$ are linear variational parameters. The nonlinear parameters
$\alpha$ and $\beta$ are again chosen and optimized stochastically.

The basis function is no longer automatically symmetric over all permutations.
It has to be symmetrized with respect to particles number 1 and 2
and therefore the symmetrization operator, $\hat{S}$, has to be included
when calculating matrix elements,\begin{equation}
\hat{S}\Phi_{12}=\left(\begin{array}{c}
N\\
2\end{array}\right)^{-1}\sum_{ij}\Phi_{ij}.\end{equation}
 This is the same type of Faddeev-like decomposition of the wave-function
as used in \cite{Ole,Ole2,Sogo}.

Fortunately, only a finite number of different terms appear in calculations
of the matrix elements, and the computational time is therefore still
independent of the number of particles. Indeed the kinetic energy
and the external field operators are fully symmetric and therefore
the explicit symmetrization of the wave-function is not needed for
their matrix elements. The matrix element for the inter-particle potentials
reduces to a finite number of terms,\begin{eqnarray}
 & \left(\begin{array}{c}
N\\
2\end{array}\right)\langle\Phi_{12}\mid\sum_{i<j}V_{ij}\hat{S}\mid\Phi_{12}\rangle=\\
 & \langle\Phi_{12}\mid\left(V_{12}+2(N-2)V_{13}+\frac{(N-2)(N-3)}{2}V_{34}\right)\mid\Phi_{12}\rangle\nonumber \\
 & +2(N-2)\langle\Phi_{12}\mid\nonumber \\
 & \left(V_{12}+V_{13}+V_{23}+(N-3)(V_{14}+V_{24}+V_{34})+\frac{(N-3)(N-4)}{2}V_{45}\right)\nonumber \\
 & \mid\Phi_{13}\rangle\nonumber \\
 & +\left(\frac{N(N-1)}{2}-1-2(N-2)\right)\langle\Phi_{12}\mid\nonumber \\
 & \left(V_{12}+4V_{13}+V_{24}+V_{34}+2(N-4)(V_{15}+V_{35})+\frac{(N-4)(N-5)}{2}V_{56}\right)\nonumber \\
 & \mid\Phi_{34}\rangle\nonumber \end{eqnarray}
 Each individual matrix element in this expression is readily calculated
using the expression (\ref{eq:AVA}) in the appendix. The structure
of the expression basically corresponds to that of \cite{Ole2} where
hyper-spherical coordinates were used instead of the Jacobi
coordinates used here. Hyper-spherical coordinates allow an easy implementation
of a powerful hyper-spheric adiabatic expansion method but, on the
other hand, do not allow an easy implementation of higher order correlations.

\subsubsection{Three-body correlations}

The three-body correlations can be accounted for by a basis function
of the form
\begin{equation}
Phi_{123}=e^{-\frac{1}{2}\alpha\rho^{2}-\frac{1}{2}\beta(\mathbf{q}_{1}-\mathbf{q}_{2})^{2}-\frac{1}{2}\gamma(\mathbf{q}_{1}-\mathbf{q}_{3})^{2}},
\end{equation}
where $\alpha$, $\beta$ and $\gamma$ are independent parameters.
The trial wave-function is then a linear combination of $\Phi_{123}$'s
with different parameters $\alpha$ and $\beta$ and $\gamma$,\begin{equation}
\Psi=\sum_{s,u,v}C_{suv}e^{-\frac{1}{2}\alpha_{s}\rho^{2}-\frac{1}{2}\beta_{u}(\mathbf{q}_{1}-\mathbf{q}_{2})^{2}-\frac{1}{2}\gamma_{v}(\mathbf{q}_{1}-\mathbf{q}_{3})^{2}},\end{equation}
where $C_{suv}$ are linear variational parameters, and where the
nonlinear parameters $\alpha$, $\beta$ and $\gamma$ are again chosen
and optimized stochastically.

This function must be explicitly symmetrized with respect to particles
1, 2, and 3. This symmetrization again results in a finite number
of different terms as it did for two-body correlations. There are
in total 34 different terms and it is therefore not practical to write
them down here. The computer program can easily catch the identical
terms and thus reduce the computational complexity down to the order
of $O(1)$, that is, independent of the number of particles.

\section{Numerical illustrations}

\subsection{The Bose system\label{sub:The-system}}

We use $^{87}\mathrm{Rb}$ condensate parameters corresponding to
fixed scattering length $a_{s}=100$~a.u. and trapping frequency
$\omega=2\pi\times77.87$ Hz, and vary the number of atoms
$N=10^{1}-10^{4}$.  In all cases, the inverse square root of the
nonlinear parameters $\beta_{k}$ and $\gamma_{k}$ are optimized
from the random value interval $[10^{-4}b_{t};10b_{t}]$ (where
$b_{t}=\sqrt{\hbar/(m\omega)}\approx23095$~a.u. is the trap length), while
for the parameters $\alpha$ the interval was $[b_{t};10^{3}b_{t}]$. In
practice only one parameter $\alpha_{0}$ was needed to achieve the chosen
accuracy goal of three digits on the interaction energy per particle.

The mean-field validity condition, $na_{s}^{3}\ll~1$, where $n$ is the
particle density, is fulfilled for all values of $N$. Therefore the
Gross-Pitaevskii results from the literature should be quite accurate
and we shall use them as the reference point. The other regime,
$na_{s}^{3}\gg1$, shall be investigated separately.

\subsection{Two-body potentials}

We consider only dilute bosonic systems where the properties 
largely depend upon the low-energy/large-distance properties of the two-body
interaction, that is the s-wave scattering length $a_{s}$. In this
regime a zero-range pseudo-potential given by a delta function,
\begin{equation}
V_{\delta}(r)=\frac{4\pi\hbar^{2}a_{s}}{m}\delta(r),\label{eq:Vd}
\end{equation}
is proven to provide within a mean-field theory a good approximation
to the energy of the system. Applying the delta-function interaction
with a Hilbert space of a beyond-mean-field theory, however, requires
an appropriate renormalization \cite{Sogo}. The physical scattering
length in (\ref{eq:Vd}) should be substituted by its first-order Born
approximation of the given finite-range potential.

We shall use the delta-function potential for calculation with the
uncorrelated hyper-radial trial wave-function.

For correlated calculations we shall use four different finite-range
potentials of the form\begin{equation}
V(r)=V_{0}e^{-r^{2}/b^{2}}+U_{0}e^{-r^{2}/c^{2}},\end{equation}
where the parameters of the potentials are specified in Table~\ref{pots}.
The first potential, marked H, is a hard repulsive core, the second,
S, is a soft repulsive core, the third, A, is an attractive well,
and the fourth, W, is a semi-realistic well with a repulsive core
and an attractive pocket. All potentials have the same scattering
length, $a_{s}=100$~a.u., and in the dilute regime should therefore
provide identical energies if correlations are appropriately included.

\begin{table}

\begin{tabular}{cccccc}
\hline 
Designation&
 $b$&
 $V_{0}$&
 $c$&
 $U_{0}$&
 $N_{b}$\tabularnewline
\hline
$\textrm{H}$ (hard) &
 $58.69$&
 $1.906\times10^{-7}$&
 $0$&
 $0$&
 $0$\tabularnewline
$\textrm{S}$ (soft) &
 $550.0$&
 $1\times10^{-11}$&
 $0$&
 $0$&
 $0$\tabularnewline
$\textrm{A}$ (attractive) &
 $10$&
 $-1.906\times10^{-7}$&
 $0$&
 $0$&
 $1$\tabularnewline
$\textrm{W}$ (well) &
 $4.4$&
 $5.566\times10^{-5}$&
 $10$&
 $-1.125\times10^{-6}$&
 $1$ \tabularnewline
\hline
\end{tabular}

\caption{The parameters (in atomic units) of the finite-range Gaussian two-body
potentials of the form $V(r)=V_{0}e^{-r^{2}/b^{2}}+U_{0}e^{-r^{2}/c^{2}}$
used in the calculations. $N_{b}$ is the number of bound states in
the potential. The $s$-wave scattering length $a_{s}$ is equal 100
a.u. for all potentials.}

\label{pots}
\end{table}

\subsection{Results}

The results are collected in Tables~\ref{tab:res1} and \ref{tab:res2},
where we show the interaction energy per particle,
$\frac{E}{N}-\frac{3}{2}\hbar\omega$ (where $E$ is the total energy of the
system), for different combinations of numbers of particles, potentials,
and trial wave-functions.  The absence of a number for the attractive and
realistic potential means that there are many strongly bound (collapsed)
states and an analog of the condensate state located in the trap does
not exist.

\begin{table}\label{tab:res1}

\begin{tabular}{|c|ccc|ccc|c|c|}
\hline 
&
\multicolumn{3}{c|}{hard-core potential}&
\multicolumn{3}{c|}{soft-core potential}&
$\delta$-function&
\tabularnewline
$N$&
1b&
2b&
3b&
1b&
2b&
3b&
1b&
GP\tabularnewline
\hline
10&
.329&
.0155&
.0154&
.0179&
.0154&
.0154&
.0154&
.0154\tabularnewline
20&
.599&
.0326&
.0325&
.0373&
.0320&
.0320&
.0320&
.0320\tabularnewline
50&
1.18&
.0832&
.0828&
.0923&
.0795&
.0794&
.0798&
.0792\tabularnewline
100&
1.83&
.165&
.164&
.177&
.153&
.153&
.153&
.151\tabularnewline
1000&
6.29&
1.32&
1.32&
1.09&
1.00&
.999&
.978&
.930\tabularnewline
5000&
13.2&
4.48&
4.47&
2.88&
2.75&
2.75&
2.64&
2.45\tabularnewline
10000&
17.8&
7.27&
7.26&
4.15&
4.02&
4.02&
3.83&
3.58\tabularnewline
\hline
\end{tabular}

\caption{The interaction energy per particle,
$\frac{E}{N}-\frac{3}{2}\hbar\omega$, where $E$ is
the total energy, for the system described in
the text. Results are given for the hard-core
(H) and soft-core (S) potential from Table~\ref{pots} with different
trial wave functions (1b -- uncorrelated, 2b -- two-body correlations,
3b -- three-body correlations) as well as for the $\delta$ -function
potential with uncorrelated wave-function. The last column shows
the Gross-Pitaevskii (mean-field) results from \cite{McKinney} and
\cite{Fabrocini}.}
\end{table}

\begin{table}\label{tab:res2}

\begin{tabular}{|c|cc|ccc|c|}
\hline 
&
\multicolumn{2}{c|}{attractive potential}&
\multicolumn{3}{c|}{realistic potential}&
\tabularnewline
$N$&
1b&
2b&
1b&
2b&
3b&
GP\tabularnewline
\hline
10&
-.0021&
.0147&
.0383&
.0154&
.0154&
.0154\tabularnewline
20&
-.0044&
.0264&
.0599&
.0320&
.0320&
.0320\tabularnewline
50&
-.0114&
.0228&
.188&
.0804&
.0802&
.0792\tabularnewline
100&
-.0233&
-.0042&
.344&
.156&
.155&
.151\tabularnewline
1000&
&
&
1.78&
1.07&
&
.930\tabularnewline
5000&
&
&
4.33&
3.27&
&
2.45\tabularnewline
10000&
&
&
6.11&
5.09&
&
3.58\tabularnewline
\hline
\end{tabular}

\caption{The same as Table~\ref{tab:res1} for the attractive (A), and
realistic (W) potentials from Table~\ref{pots}. For larger number of
particles and higher correlations the potentials produce a large number
of strongly bound (collapsed) states and thus no condensate state could
have been traced.}
\end{table}

\subsubsection{Uncorrelated wave-function}

\begin{figure}\label{fig:1b}
\input{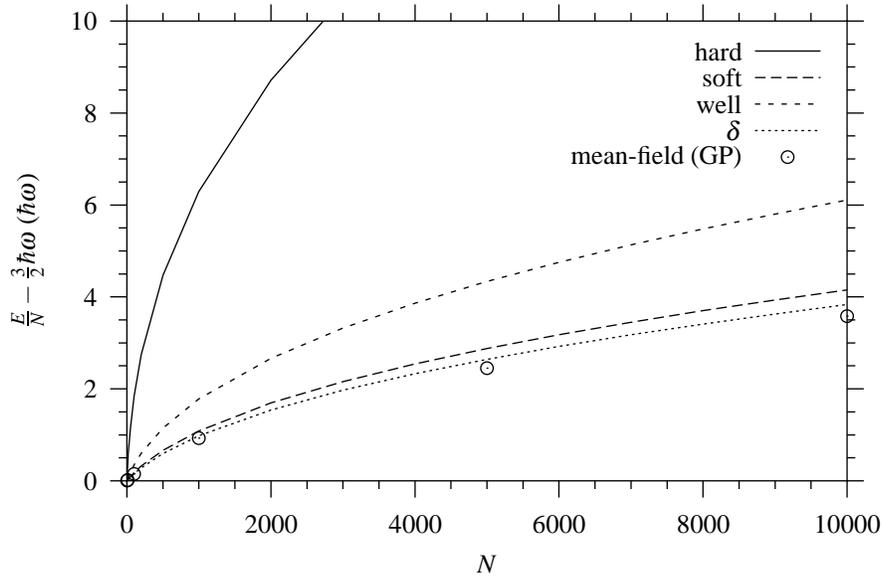}
\caption{Interaction energy per particle as function of the number of
particles $N$ for the uncorrelated trial wave-function.}
\end{figure}

The results for different potentials with the uncorrelated hyper-radial trial
wave-function are given in Tables~\ref{tab:res1}-\ref{tab:res2} and
also represented on Fig.\ref{fig:1b}.

Importantly, the combination of delta-function pseudo-potential with
the uncorrelated hyper-radial wave-function give results within a
few per cent of the mean-field theory. The pseudo-potential
therefore seems to be equally well suited for both mean-field and
hyper-radial approximations.

One can use this very fast uncorrelated pseudo-potential approximation
to a great effect as a tool to optimize the parameters of the Gaussians
to be used in the more demanding correlated calculations with finite-range
potentials.

The finite-range potentials show large deviations since the uncorrelated
wave-function is not suited for them. The hard-core potential, as
could be expected, is especially bad for the uncorrelated wave-function.
The attractive potential produces for larger number of particles a
strongly bound (collapsed) ground-state and is therefore not shown
on the figure.

\subsubsection{Two-body correlations}

\begin{figure}\label{fig:2b}
\input{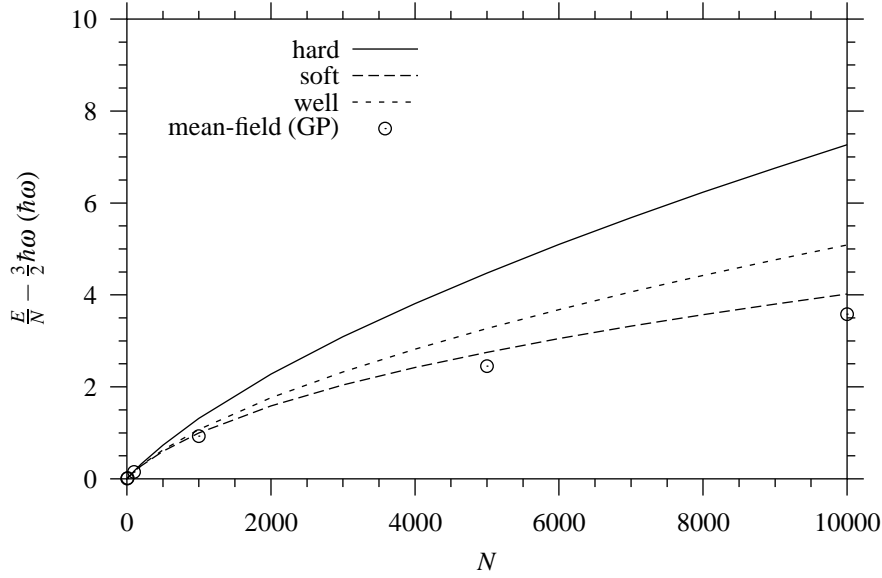}
\caption{Interaction energy per particle as function of the number of
particles $N$ for the trial wave-function with two-body correlations.}
\end{figure}

The results with the two-body correlated trial wave-function are
given in Tables~\ref{tab:res1}-\ref{tab:res2} and also represented on
Fig.~\ref{fig:2b}.

Apparently, inclusion of two-body correlations dramatically improves
the results. This seems to support the assertion in \cite{Ole,Ole2,Sogo}
that the two-body correlations are of the utmost importance for the
dilute gases.

The hard-core potential, although doing much better with the two-body
correlated wave-function, is still the farthest off especially for
large number of particles. The soft-core potential on the other hand
is now very close to the mean-field results.

\subsubsection{Three-body correlations}

We do not show a separate figure for the three-body correlations as
they turn out not to produce large effects on the energies apart from
potentials with attraction, where the three-body correlations quite
expectedly straight away lead to a large number of strongly bound
(collapsed) states.

Thus, for model repulsive finite-range potentials and dilute systems
the three-body correlations are of much less importance that two-body
correlations.

\section{Conclusion}

We have introduced a new approach, based on correlated Gaussian method,
to investigate dilute Bose systems. The approach allows to include
consecutively correlations of different orders in the trial wave-function.
In its lowest (uncorrelated) order with zero-range pseudo-potentials
the approach is comparable to the mean-field (Gross-Pitaevskii) theory.

We have performed an exploratory numerical investigation of two- and
three-body correlation in a dilute Bose system with different number
of particles and different finite-range potentials. For the condensate
state the two-body correlations turn out to be by far the most important
and suffice to provide a quantitative description of the system with
soft-core potentials.

\begin{theacknowledgments}
H.~H. Sørensen would like to thank Christoffer Dam Bruun for many
interesting discussions and valuable assistance during implementation
of the method.
\end{theacknowledgments}

\section{Appendix: correlated Gaussian method}

The trial wave-function is represented as a linear combination of
correlated Gaussians,$\left|A\right\rangle $, which have the form
\begin{equation}
A=\exp\left(-\frac{1}{2}\sum_{i,j=1}^{N}(\mathbf{x}_{i}\cdot\mathbf{x}_{j})A_{ij}\right)\equiv\exp\left(-\frac{1}{2}\mathbf{x}^{T}A\mathbf{x}\right)\;,
\end{equation}
where $A$ is a positively definite symmetric matrix and $\mathbf{x}$
is a set of (scaled Jacobi) coordinates. Correlated Gaussians form
a full basis since any square-integrable function can be represented as
a linear combination of Gaussians with arbitrary precision. The elements
of the parameter matrix $A$ can be optimize using the stochastic method
\cite{SVM}.

The important matrix elements which are used in the calculations are
the overlap of two Gaussians,
\begin{equation}
\langle A|A'\rangle=\left(\frac{(2\pi)^{N}}{\det(A+A')}\right)^{3/2},
\end{equation}
the matrix element of the kinetic energy operator,
\begin{equation}
\langle A|-\frac{\hbar^{2}}{2m}\sum_{i}\frac{\partial^{2}}{\partial\mathbf{x}_{i}^{2}}|A'\rangle
=\frac{\hbar^{2}}{2m}3\mathrm{tr}\left((A+A')^{-1}AA'\right)\langle A\mid A'\rangle,
\end{equation}
and the matrix element of the two-body potential
$V(\mathbf{r}_{i}-\mathbf{r}_{j})$,
\begin{equation}\label{eq:AVA}
\langle A\mid V(\mathbf{r}_{i}-\mathbf{r}_{j})\mid A'\rangle=\int_{-\infty}^{+\infty}d^{3}rV(\mathbf{r})\langle A\mid\delta(b_{ij}^{T}\mathbf{x}-\mathbf{r})\mid A'\rangle
=G_{c_{ij}}[V]\langle A\mid A'\rangle,
\end{equation}
 where $\mathbf{r}_{i}-\mathbf{r}_{j}=b_{ij}^{T}\mathbf{x}$, $c_{ij}^{-1}=b_{ij}^{T}(A+A')^{-1}b_{ij}$,
and $G_{c}[V]$ is the Gaussian transform of the potential\begin{equation}
G_{c}[V]=\left(\frac{c}{2\pi}\right)^{3/2}\int d^{3}rV(\mathbf{r})e^{-\frac{1}{2}cr^{2}}.\end{equation}

Other useful integrals
\begin{equation}
\langle A \mid \mathbf{x}^{T}B\mathbf{x} \mid A' \rangle
= 3\mathrm{tr}\left((A+A')^{-1}B\right) \langle A \mid A' \rangle \;;
\end{equation}
\begin{equation}
\langle A \mid \delta(\mathbf{b}^{T}\mathbf{x}-\mathbf{q}) \mid A' \rangle
= \left( \frac{\beta}{2\pi} \right)^{3/2} e^{-\frac{1}{2}\beta r^{2}}
\langle A \mid A' \rangle \;,\;\mathrm{where}\;
\beta^{-1}=\mathbf{b}^{T}(A+A')^{-1}\mathbf{b} \;;
\end{equation}
\begin{equation}
G_c[\frac{1}{r}] = 2\sqrt{\frac{c}{2\pi}} \;;
\end{equation}
\begin{equation}
G_c[\delta(r)] = \left( \frac{c}{2\pi} \right)^{3/2} \;;
\end{equation}
\begin{equation}
G_c[e^{-\frac{1}{2}kr^2}] = \left( \frac{c}{c+k} \right)^{3/2} \;.
\end{equation}

\bibliographystyle{aipproc}

\end{document}